\documentclass[journal]{IEEEtran}

\usepackage{url}

\usepackage{graphicx}

\usepackage{amsfonts}
\usepackage{amsmath}
\usepackage{cite}
\usepackage{float}
\usepackage{theorem}
\usepackage{amssymb}
\usepackage{booktabs}
\usepackage{tikz}
\usetikzlibrary{shapes,arrows}
\newtheorem{theorem}{Theorem}
\usepackage{subfig}
\usepackage{makecell}
\usepackage{enumitem}
\usepackage{multirow}
\usepackage{balance}

\newcolumntype{C}[1]{>{\centering\arraybackslash}m{#1}}
\usetikzlibrary{positioning, fit, arrows.meta, shapes,intersections,calc}
\begin{document}
	
	\title{Joint Forecasting and Interpolation of Graph Signals Using Deep Learning}

	\author{ Gabriela Lewenfus, Wallace Alves Martins, Symeon Chatzinotas, and Bj\"orn Ottersten
		\thanks{Ms. Lewenfus is with the Federal University of Rio de Janeiro (UFRJ), (e-mail: gabriela.lewenfus@gmail.com);
			Prof. Martins is with UFRJ (on leave) and with the University of Luxembourg (as Research Associate), (e-mail: wallace.martins@smt.ufrj.br);  Prof. Chatzinotas is with the University of Luxembourg (SnT), (e-mail: symeon.chatzinotas@uni.lu); and Prof. Ottersten is with the University of Luxembourg (SnT), (e-mail: bjorn.ottersten@uni.lu)}
			\thanks{This work was financed in part by the ERC project AGNOSTIC and by the Brazilian research agencies CNPq and Faperj.}
			\thanks{© 2020 IEEE.  Personal use of this material is permitted.  Permission from IEEE must be obtained for all other uses, in any current or future media, including reprinting/republishing this material for advertising or promotional purposes, creating new collective works, for resale or redistribution to servers or lists, or reuse of any copyrighted component of this work in other works.}}

	\markboth{}
	{Shell \MakeLowercase{\textit{et al.}}: Bare Demo of IEEEtran.cls for IEEE Journals}
	\maketitle
	
	\begin{abstract}
	We tackle the problem of forecasting  network-signal snapshots using past signal measurements acquired by a subset of network nodes. This task can be seen as a combination of multivariate time-series prediction and graph-signal interpolation. This is a fundamental problem for many applications wherein deploying a high granularity network is impractical. Our solution combines recurrent neural networks with frequency-analysis tools from graph signal processing, and assumes that data is sufficiently smooth with respect to the underlying graph. The proposed approach outperforms   state-of-the-art deep learning techniques, especially when only a small fraction of the graph signals is accessible, considering two distinct real world datasets: temperatures in the US and speed flow in Seattle. The results also indicate that our method can  handle noisy signals and missing data, making it suitable to many practical applications.
	\end{abstract}
	
	\IEEEpeerreviewmaketitle
	
	\section{Introduction}
	\IEEEPARstart{S}{patiotemporal} (ST) prediction is a fundamental abstract problem featuring in many practical applications, including climate analyses~\cite{Salman2016}, transportation management~\cite{Lv2015}, neuroscience~\cite{Smith2011}, electricity markets\cite{Li2017}, and several geographical phenomenon analyses~\cite{Racah2017}. 
	 The temperature in a city, for instance,  is influenced by its location, by the season, and even by the hour of the day. Another example of data with ST dependencies is the traffic state of a road, since it is influenced by adjacent roads and also by the hour of the day. ST prediction boils down to \emph{forecasting} (temporal prediction) and \emph{interpolation} (spatial prediction). The former refers to predicting some physical phenomenon using historical data acquired by a network of spatially-distributed sensors. The latter refers to predicting the phenomenon with a higher spatial resolution. In this context, ST data can be seen as a network signal in which a time series is associated with each network element; the dynamics (time-domain evolution) of the time series depends on the network structure (spatial domain), rather than on the isolated network elements only. The interpolation is useful to generate a denser (virtual) network. 
	 
	  Classical predictive models generally assume independence of data samples and disregard relevant spatial information~\cite{williams2003modeling},~\cite{Cai2019}. Vector autoregressive (VAR)~\cite{chandra2009predictions}, a statistical multivariate model, and machine learning (ML) approaches, such as support vector regression (SVR)~\cite{kramer2011short} and random forest regression~\cite{leshem2007traffic}, can achieve higher accuracy  than  classical predictive models; yet, they fail to fully capture spatial relations. 
	  More recently, some progress has been made by applying neural networks\footnote{In this paper the word ``network" can refer to a neural network in the context of deep learning or a physical network that is represented by a graph.} (NNs) to predict ST data~\cite{huang2014deep,Lv2015,jia2016traffic,Salman2016,Salman2018,Liang2019}.
	  NNs have the capacity of not only mapping an input  data to an output, but also of learning a useful representation to improve the mapping accuracy~\cite{goodfellow2016deep}.  Nonetheless, the fully-connected architecture of these NNs fails to extract  simultaneous spatial and temporal features from data.  	
	
	In order to learn spatial information from these multivariate time series, some works have combined convolutional NNs (CNNs) with recurrent NNs (RNNs), such as long short-term memory (LSTM)~\cite{Shi2015a,Yu2017,Huang2018b,Wu2018,Li2019a}. However, CNNs are restricted to grid-like uniformly structured data, such as images and videos. To overcome this issue, and inspired by graph\footnote{Graphs are mathematical structures able to represent rather general datasets, including ST data with irregular domains, as in sensor networks.} signal processing (GSP), some works have developed convolutions on graph-structured data (graph signals)~\cite{Henaff2015,Defferrard2016a,Levie2019,Kipf2019,Ata2018}, which have been used in combination with either RNN, time convolution, and/or attention mechanisms to make predictions in a variety of applications. These works are summarized in TABLE~\ref{tab:st_summary}. 

\begin{table}[htb!]
    \centering
    \caption{Summary of recent works that use deep learning to predict ST data. First column defines the application. Second and third columns refer to the spatial and temporal  techniques employed, respectively. ``Conv.'' means temporal convolution; AE means auto encoder; RBM means restricted Boltzman machine; ``other'' encompasses other predictive strategies, such as attention mechanisms}

{\def\arraystretch{1.5}\tabcolsep=8pt
\begin{tabular}{|c|l|l|p{0.18\textwidth}|}
\hline
Application & GSP& Temporal &Paper\\ \hline
\multirow{6}{*}{traffic}&\multirow{3}{*}{Yes}& RNN & 

\cite{Ata2018},~\cite{zhao2018temporal,Zhang2018a,wang2018,Geng2019,chen2019multi,Cui2019}\\\cline{3-4}
&&Conv.&
\cite{Yu2018,wang2018dynamic,Wu2019a,Lee2019,diao2019dynamic,fang2019gstnet}\\\cline{3-4}
&&other&
\cite{Song2020},~\cite{guo2019attention}\\\cline{2-4}
&\multirow{2}{*}{No}& RNN & 
\cite{Yu2017},~\cite{Li2019a},~\cite{Cui2018,Wang2018a,Liao2018a,Liao2018,Han2019},\\ \cline{3-4}
&&other&\cite{Lv2015}~,\cite{Wu2018},~\cite{Lv2018,Lin2019,Yang2017,Zhang2019a,Ouyang2016}\\\hline

\multirow{3}{*}{wind}&\multirow{1}{*}{Yes}& RNN &\cite{Khodayar2019} \\\cline{2-4}
&\multirow{2}{*}{No}& RNN & \cite{Liang2019}\\\cline{3-4}
&& Other & \cite{Khodayar2017,Zhu2018,Zhang2015}\\\hline

\multirow{3}{*}{meteorological}&\multirow{1}{*}{Yes}& AE & \cite{Khodayar2020}\\\cline{2-4}
&\multirow{2}{*}{No}& RNN & \cite{Shi2015a}\\\cline{3-4}
&& Other & \cite{Rasp2018a,Racah2017}\\\hline

\multirow{1}{*}{body-motion}&\multirow{1}{*}{Yes}& RNN & \cite{Li2018,wu2019spatial,Si2019}\\\cline{1-4}%

\multirow{2}{*}{neuroscience}&\multirow{1}{*}{Yes}& Conv. & \cite{gadgil2020spatio}\\\cline{2-4}
&\multirow{1}{*}{No}& RBM & \cite{Huang2016a}\\\hline
\multirow{1}{*}{semantic}&\multirow{1}{*}{Yes}& RNN & \cite{Li2016a}\\\cline{1-4}%
\end{tabular}

}

    \label{tab:st_summary}
\end{table}	
	
	
	GSP theory has been applied to analyze/process many irregularly structured datasets in several  applications~\cite{Shuman2013,Angeles2018}. An import task addressed by GSP is interpolation on graphs, i.e., (spatially) predicting the signals on a subset of graph nodes based on known signal values from other nodes~\cite{Narang2013a}. In general, graph interpolation is based on local or global approaches. Local methods, such as $k$-nearest neighbors ($k$-NNs)~\cite{Chen2009}, compute the unknown signal values in a set of network nodes using values from their closest neighbors, being computationally efficient. Global methods, on the other hand, interpolate the unknown signal values at once and can provide better results by taking the entire network into account at the expensive of a higher computational burden~\cite{Narang2013a,Segarra2015}. Many GSP-based interpolation techniques have been proposed~\cite{Gavish2010,Pesenson2011,Narang2013a,Anis2014,Chen2014,Chen2015,Naranga,Chen2015,Spelta2018a,spelta2020normalized}. Due to the irregular structure of graph signals, the interpolation problem may become ill-conditioned, calling for efficient selection strategies for obtaining optimal sampling sets~\cite{Anis2016a,Chamon2018}. 	
	In fact, the problem of interpolating a graph signal (GS) can also be addressed as  semi-supervised classification~\cite{DavidIShumanMohammadJavadFaraji2011,Gadde2014,Chen2016,Parisot2018,Cheung2018}  or regression~\cite{Huang2017,Huang2018,Manohar2018} tasks. More recently, deep learning (DL) solutions have also been developed~\cite{Kipf2019,Xu2019a}. 
	
	In many applications, affording to work with large networks may be impractical; for example, placing many electrodes at once in the human cortex may be unfeasible. Installation and maintenance costs of devices can also limit the number of sensors deployed in a network~\cite{Manohar2018}. Thus, developing a predictive model capable of forecasting (temporal prediction) and interpolating (spatial prediction) time-varying signals defined on graph nodes can be of great applicability. This problem can be regarded as a semi-supervised task, since only part of the nodes are available for training. Other works have addressed this problem: in~\cite{Romero2017} the graph is extended to incorporate the time dimension and a kernel-based algorithm is used for prediction;  this approach, therefore, relies on the assumption of smoothness in the time domain, which is not reasonable for many applications, such as traffic flow prediction. In~\cite{Khodayar2019}, the ST wind speed model is evaluated in a semi-supervised framework in which only part of the nodes are used for training the model, while interpolation is performed only in test phase. Therefore, the parameters learned during the training phase do not take into account the interpolation aspect.
	
	Two straightforward solutions to deal with the problem of forecasting and interpolating sampled GSs are:\footnote{We can regard a GS that will be interpolated as a sampled version of a GS defined over a denser set of nodes belonging to a virtual graph.} (i) applying a forecasting model to the input GS and then interpolating the output; or (ii) interpolating the sampled GS and then feeding it to the forecasting model. These solutions tackle the ST prediction task  separately and may fail to capture the inherent coupling between time and space domains. 
	In this paper, a graph-based NN architecture is proposed to handle ST correlations  by employing GSP in conjunction with a gated-recurrent unit (GRU). Thus, we address the  inherent nature of ST data by jointly forecasting and interpolating the underlying network signals. A global interpolation approach is adopted as it provides accurate results when the signal is smooth in the GSP sense, whereas an RNN forecasting model is adopted given its prior success in network prediction. Herein, not only the sampled GS is input to a predictive model but also its spectral components, which carry spatial information on the  underlying graph. The major contribution of our proposed model is, therefore, the ability to learn ST features by observing a few nodes of the entire graph.

Considering the proposed learning framework, we introduce four possible classes of problems: 
\begin{itemize}
    \item supervised applications, where the labels of all nodes are available for training but only a fixed subset of graph nodes can be used as input to the model in the test phase;
    \item semi-supervised application, wherein only data associated with a subset of nodes are available for training and computing gradients;
    \item noise-corrupted application, in which all nodes are available during the entire process, but additive noise corrupts the network signals;
    \item missing-value application, where a time-varying fraction of nodes are available for testing, but all nodes can be used for training. 
\end{itemize}
The proposed approach achieves  the best results in most tested scenarios related to the aforementioned applications, as compared to DL-based benchmarks.  
	
	The paper is organized as follows: Section~\ref{sec:gsp} presents some fundamental aspects of GSP, focusing on the sampling theory that will be used to build the interpolation module of the proposed learning model.   Section~\ref{sec:methods} describes the new learning framework. Section ~\ref{sec:applications} describes four classes of applications that can benefit from the proposal.  Section~\ref{sec:results} presents the numerical results and related discussions. Section~\ref{sec:conclusion} contains the concluding remarks of the paper.

	\section{GSP Background}
	\label{sec:gsp}
	Let $\mathcal{G} \triangleq (\mathcal{V},\mathcal{E}, \mathbf{A})$ be a weighted undirected and connected graph, where $\mathcal{V} \triangleq \{v_1, \ldots, v_N\}$ is the set of $N$ nodes, $\mathcal{E}$ is the set of edges, and $\mathbf{A}$ is the $N\times N$ adjacency matrix containing edge weights $A_{mn}$. The adjacency matrix can be a similarity matrix or be built based on prior information, such as nodes' locations in the physical network.
	  A GS is  a real-valued scalar function $x: \mathcal{V} \to \mathbb{R}$ taking values on the graph nodes, and it will be represented by the $N$-dimensional vector $\mathbf{x}$ with entries $\left[ \mathbf{x}\right]_n = x_n = x(v_n)$.
	 
	The diagonal matrix $\mathbf{D} \in \mathbb{R}^{N\times N}$ is the degree matrix in which $D_{nn} = \sum_m A_{nm}$ measures the connectivity degree of each node. 
	
	  Most of the graph convolutions in the literature are based on the Laplacian matrix $\mathbf{L} \triangleq \mathbf{D}-\mathbf{A}$, which is a symmetric semi-definite matrix, for a symmetric adjacency ${\bf A}$.  Let $\mathbf{U}$ be the matrix of orthonormal eigenvectors of $\mathbf{L}$. The graph Fourier transform (GFT) of the GS  $\mathbf{x}$ is $\hat{\mathbf{x}} \triangleq \mathbf{U}^{\rm T}\mathbf{x}$ and the graph frequencies are considered as the eigenvalues of $\mathbf{L}$, $\lambda_1,...,\lambda_N\geq 0$. The eigenvectors of the adjacency matrix can also be used as the Fourier basis~\cite{Sandryhaila2013a}. For the reader's convenience, TABLE~\ref{tab:notations} contains the main notations that will be used in this work.
	 
	\subsection{GS Sampling}\label{se:sampling}
	Let $\mathcal{S} \triangleq \{s_1, ...,s_M \}  \subset{\mathcal{V}}$  be a subset of nodes with $M \leq N$ nodes; the vector of measurements $\mathbf{x}_{\mathcal{S}} \in \mathbb{R}^M$ is given by $\mathbf{x}_{\mathcal{S}} = \boldsymbol{\Psi}_{\mathcal{S}}\mathbf{x}$, where the sampling operator
	\begin{align}
	\left[\boldsymbol{\Psi}_{\mathcal{S}}\right]_{mn} = \left\{ \begin{array}{cc}
	1, & \textrm{if\;} v_n = s_m \\
	0, & {\rm otherwise}
	\end{array}\right.
	\end{align} 
	 selects from $\mathcal{V}$ the nodes in $\mathcal{S}$.
	  The interpolation operator $\boldsymbol{\Phi}_{\mathcal{S}}$ is an $N\times M$ matrix such that the recovered signal is 
	 	$\tilde{\mathbf{x}}=\boldsymbol{\Phi}_{\mathcal{S}}\boldsymbol{\Psi}_{\mathcal{S}}\mathbf{x}$. If $\tilde{\mathbf{x}} = \mathbf{x}$, the pair of sampling and interpolation operators $(\boldsymbol{\Phi}_{\mathcal{S}},\boldsymbol{\Psi}_{\mathcal{S}})$ can perfectly recover the signal $\mathbf{x}$ from its sampled version. As the rank of $\boldsymbol{\Phi}_{\mathcal{S}}\boldsymbol{\Psi}_{\mathcal{S}}$ is smaller or equal to $M$, this is not possible for all $\mathbf{x}\in \mathbb{R}^N$ when $M<N$. However, perfect reconstruction can be achieved for a class of bandlimited GS.
	 	
	 The GS $\mathbf{x}_{\rm b}$ is said $\mathcal{F}$-bandlimited if $\left[\hat{\bf x}_{\rm b}\right]_n = 0 \; \forall n $ such that $\lambda_n \not\in \mathcal{F}\subset \{\lambda_1,\ldots, \lambda_N\} $, that is, the frequency content of $\mathbf{x}_{\rm b}$ is restricted to the set of frequencies $\mathcal{F}$. Some works also restrict the support of the frequency content and consider that a GS $\mathbf{x}_{\rm b}$ is $\omega$-bandlimited if  $\left[\hat{\bf x}_{\rm b}\right]_n = 0 \; \forall n$ such that $\lambda_n >\omega $\cite{Narang2011}. In this paper, a bandlimited signal is a sparse vector in the GFT domain. The following theorem guarantees the perfect reconstruction of an $\mathcal{F}$-bandlimited GS for some sampling sets.
	 \begin{theorem}(\cite{Chen2015},\cite{Pesenson2011})
	 	If the sampling operator $\boldsymbol{\Psi}_{\mathcal{S}}$ satisfies 
	 	\begin{align}
	 	{\rm rank}(\boldsymbol{\Psi}_{\mathcal{S}}\mathbf{U}_{:,\mathcal{F}}) = |\mathcal{F}|=K,	\label{eq:sampling_condition}
	 	\end{align}	
	 	then $\mathbf{x}_{\rm b} = \boldsymbol{\Phi}_{\mathcal{S}}\boldsymbol{\Psi}_{\mathcal{S}}\mathbf{x}_{\rm b}$ as long as  $\boldsymbol{\Phi}_{\mathcal{S}}=\mathbf{U}_{:,\mathcal{F}}\boldsymbol{\Sigma}$, with $\boldsymbol{\Sigma}$ satisfying $\boldsymbol{\Sigma}\boldsymbol{\Psi}_{\mathcal{S}}\mathbf{U}_{:,\mathcal{F}}=\mathbf{I}_K$ and $\mathbf{U}_{:,\mathcal{F}}$ a submatrix of $\mathbf{U}$ with columns restricted to the indices associated with the frequencies in $\mathcal{F}$.
	 \end{theorem}
	 
	The condition in \eqref{eq:sampling_condition} is also equivalent to 
	\begin{align}
	{\rm SV}_{\max}(\mathbf{U}_{\overline{\mathcal{S}},\mathcal{F}})\leq 1, \label{eq:sampling_condition2}
	\end{align}
	where ${\rm SV}_{\max}(.)$ stands for the largest singular value~\cite{Lorenzo2018} and $\overline{\mathcal{S}} = \mathcal{V}/\mathcal{S}$. This means that no $\mathcal{F}$-bandlimited signal over the graph $\mathcal{G}$ is supported on $\overline{\mathcal{S}}$.	
	 
	  In order to have $\boldsymbol{\Sigma}\boldsymbol{\Psi}_{\mathcal{S}}\mathbf{V}_{:,\mathcal{F}}=\mathbf{I}_K$ we must have $M\geq K$, since ${\rm rank}(\mathbf{U}_{:,\mathcal{F}}) = K$. If $M\geq K$, $\boldsymbol{\Sigma}$ is the pseudo-inverse of $\boldsymbol{\Psi}_{\mathcal{S}}\mathbf{U}_{:,\mathcal{F}}$ and the interpolation operator is
	 \begin{align}
	 \boldsymbol{\Phi} = \mathbf{U}_{:,\mathcal{F}}(\mathbf{U}_{:,\mathcal{F}}^{\rm T}\boldsymbol{\Psi}_{\mathcal{S}}\mathbf{U}_{:,\mathcal{F}})^{-1}\mathbf{U}_{\mathcal{S},\mathcal{F}}^{\rm T}\,. \label{eq:recovery}
	 \end{align}
	 Since $\mathbf{U}$ is non-singular, there is always at least a subset $\mathcal{S}$ such that the condition in \eqref{eq:sampling_condition} is satisfied. Nonetheless, for many choices of $\mathcal{S}$, $\boldsymbol{\Psi}_{\mathcal{S}}\mathbf{U}_{:,\mathcal{F}}$ can be full rank but ill-conditioned, leading to large reconstruction errors, especially in the presence of noisy measurements or in the case of approximately bandlimited GS. To overcome this issue, optimal sampling strategies, in the sense of minimizing reconstruction error, can be employed~\cite{Chen2015}. Note that $\boldsymbol{ \Phi}$ depends on both $\mathcal{S}$ and $\mathcal{F}$, but this dependence is omitted in order to simplify the notation. 
	 
	 \subsection{Approximately Bandlimited GS}
	 
	 In practice, most GSs are only approximately bandlimited\cite{Chen2016}.
	 A GS is approximately $(\mathcal{F},\epsilon)$-bandlimited if~\cite{Lorenzo2018} 
	 \begin{align}
	 \mathbf{x} = \mathbf{x}_{\rm b} + \boldsymbol{\eta},
	 \end{align}
	 where $\mathbf{x}_{\rm b}$ is an $\mathcal{F}$-bandlimited GS and $\boldsymbol{\eta}$ is an $\overline{\mathcal{F}}$-bandlimited GS such that $\|\boldsymbol{\eta}\|_2 <\epsilon$. If signal $\mathbf{x}$ is sampled on the subset $\mathcal{S}$ and recovered by the interpolator in \eqref{eq:recovery}, the error energy of the reconstructed signal is upper bounded by
	 \begin{align}
	 \|\tilde{\mathbf{x}}-\mathbf{x}\|_2 \leq \frac{\|\boldsymbol{\eta}\|_2}{{\rm cos}(\theta_{\mathcal{S},\mathcal{F}})},\label{eq:error_bound}
	 \end{align}
	 where $\theta_{\mathcal{S},\mathcal{F}}$ is the maximum angle between the subspace of signals supported on $\mathcal{S}$ and the subspace of $\mathcal{F}$-bandlimited GS. It can be shown that ${\rm cos}(\theta_{\mathcal{S},\mathcal{F}}) = {\rm SV}_{\min}(\boldsymbol{\Psi}_{\mathcal{S}}\mathbf{U}_{:,\mathcal{F}})$; therefore, in order to minimize the upper bound of the reconstruction error in~\eqref{eq:error_bound}, the  set $\mathcal{S}$ should  maximize ${\rm SV}_{\min}(\boldsymbol{\Psi}_{\mathcal{S}}\mathbf{U}_{:,\mathcal{F}})$. Finding the optimal set $\mathcal{S}$ is a combinatorial optimization problem that can  require an exhaustive search in all possible subsets of $\mathcal{V}$ with size $M$. A suboptimal solution can be obtained by the greedy search in~\cite[Algorithm~1]{Chen2015}. 
	 
	 \begin{table}[H]
	 \centering
	 \caption{Notations}
	 \label{tab:notations}
	 	{\def\arraystretch{1.5}\tabcolsep=7pt
	 	\begin{tabular}{|l|p{0.3\textwidth}|}
	 		\hline
	 		Notation & Definition\\ \hline
	 		$\mathcal{G}$ & graph\\\hline
	 		$\mathcal{V}$ & entire set of graph nodes\\\hline
	 		$\mathcal{S}$ & subset of graph nodes\\\hline
	 		$\mathcal{F}$ & subset of graph spectrum\\\hline
	 		$\overline{\mathcal{S}}$ & the complement of the set $\mathcal{S}$ \\\hline
	 		$\mathbf{L}$ & Laplacian matrix\\\hline
	 		$\mathbf{U}$ & matrix of Laplacian eigenvectors\\\hline
	 		$\mathbf{U}_{:,\mathcal{F}}$ & submatrix of $\mathbf{U}$ with columns in the set $\mathcal{F}$\\\hline
	 		$\mathbf{U}_{\mathcal{S},\mathcal{F}}$ & submatrix of $\mathbf{U}$ with columns in the set $\mathcal{F}$ and rows in $\mathcal{S}$\\\hline
	 		$\boldsymbol{\Psi}_{\mathcal{S}}$ & sampling operator $\mathcal{V} \to \mathcal{S}$\\\hline
	 		$\boldsymbol{\Phi}_{\mathcal{S}}$ & interpolation operator $\mathcal{S} \to \mathcal{V}$\\\hline
	 		$\mathbf{x}_{\mathcal{S}}$ & signal $\mathbf{x}$ restricted to the set $\mathcal{S}$\\\hline
	 		$\hat{\mathbf{x}}_{\mathcal{F}}$ & the frequency content of the GS $\mathbf{x}$ restricted to the set $\mathcal{F}$\\\hline
	 	\end{tabular}}
	 \end{table}
	
	\section{Joint Forecasting and Interpolation of GSs}
	\label{sec:methods}
	In this section, we propose an ST neural network to jointly interpolate graph nodes and forecast future signal values. More specifically, the task is to  predict the future state $\mathbf{x}^{t+p}$ of a network given the history $\mathbf{X}_{\mathcal{S}}^{t} = \{\mathbf{x}_{\mathcal{S}}^{t},...,\mathbf{x}_{\mathcal{S}}^{t-\tau+1}\}$.\footnote{The GS at timestamp $t$ is denoted by bold lowercase letter, $\mathbf{x}^t$, whereas the history set containing the sampled GSs in previous timestamps is denoted by bold capital letter, $\mathbf{X}_{\mathcal{S}}^{t}$.} Thus, the input signal is a GS composed by $M$ nodes and the output GS is a network-signal snapshot composed by $N\geq M$ nodes. For now, to describe the learning model's architecture, we shall assume $p=1$.
	
\subsection{Gated-Recurrent Unit}\label{sec:gru}
\begin{figure}
		\centering
	  \begin{tikzpicture}[
    font=\small,
    >=LaTeX,
    cell/.style={
        rectangle, 
        rounded corners=3mm, 
        draw,
        very thick,
        },
        gate/.style={
        rectangle, 
        draw=blue,
        dashed,
        thick,
        },
    operator/.style={
        circle,
        draw,
        inner sep=-0.5pt,
        minimum height =.5cm,
        },
    state/.style={inner sep=1pt},
    branch/.style={
        circle,
        fill,
        inner sep=0pt,
        },
    ct/.style={
        circle,
        draw,
        line width = .75pt,
        minimum width=1cm,
        inner sep=1pt,
        },
    gt/.style={
        rectangle,
        draw,
        minimum width=4mm,
        minimum height=3mm,
        inner sep=1pt
        },
    ArrowC1/.style={
        rounded corners=.25cm,
        thick,
        },
    Arrowy/.style={
        rounded corners=.5cm,
        thick,
        },
    ]


    \node[operator] (legend_x) at (-1,-3) {$\times$};
    \node (elementwise) at (-1,-3.7) {element-wise};
     \node (multiplication) at (-1,-4.1) {multiplication};
     \node[operator] (legend_s) at (1.5,-3) {$+$};
     \node (elementwise2) at (1.5,-3.7) {element-wise};
    \node (add) at (1.5,-4.1) {addition};
    \node [cell, minimum height =4cm, minimum width=5.5cm] at (0,0){} ;

    \node [gt] (sigma2) at (-0.5,-0.6) {$\sigma$};
    \node [gt] (sigma1) at (0.7,-0.6) {$\sigma$};
    \node [gt, minimum width=1cm] (tanh) at (2.1,-0.8) {Tanh};
   
    \node [branch] (b1) at (-1.5,-1) {};
    \node [branch] (b2) at (0,-1.5) {};
    
    \node [branch] (b4) at (-2,-0.5) {};
    
    \node  (i1) at (-1.2,-0.99) {};
    \node  (i2) at (-1.2,-1.01) {};
    \node [operator] (mux1) at (0.7,1.5) {$\times$};
    \node [operator] (add1) at (2.1,1.5) {+};
    \node [operator] (mux2) at (-1.2,0.2) {$\times$};
    \node [operator] (mux3) at (2.1,0.2) {$\times$};
    \node [operator] (minus) at (0.7,0.6) {$-1$};

    \node (h) at (-4,1.5) {$\mathbf{h}^{t-1}$};
    \node[state] (x) at (-2,-2.4) {$\mathbf{x}^{t}$};
    \node[state] (ct) at (1.8,-0.4) {$\mathbf{c}^{t}$};
    \node[state] (zt) at (1,-0.2) {$\mathbf{z}^{t}$};
    \node[state] (rt) at (-0.2,-0.2) {$\mathbf{r}^{t}$};
    \node(y) at (4,1.5) {$\mathbf{x}^{t+1}$};
    \node (h2) at (2.1,3) {$\mathbf{h}^{t}$};

    \draw [->,ArrowC1] (h) -- (mux1) -- (add1) -> (y);

    
    \draw [ArrowC1] (x -| b1) ++(-0.5,0.15) |- (b1); 
    \draw [ArrowC1] (x -| b2) ++(-2,0.15) |- (b2); 
     
    \draw [-,ArrowC1] (h)-| (b4); 
    \draw [ArrowC1] (mux2) -- (i1); 
    \draw [ArrowC1] (i2) |- (b2); 
    \draw [-,ArrowC1] (b4) |- (b1); 
    \draw [ArrowC1] (b1 -| sigma1)++(-0.5,0) -| (sigma1); 
    \draw [ArrowC1] (b1 -| sigma2)++(-0.5,0) -| (sigma2);
    \draw [ArrowC1] (b2 -| tanh)++(-0.5,0) -| (tanh);
    \draw [ArrowC1] (h -| mux2)++(-0.5,0) -| (mux2);

    \draw [->, Arrowy] (sigma1) -- (minus);
    \draw [->, ArrowC1] (sigma2) |- (mux2);
    \draw [->, ArrowC1] (sigma1 |- mux3)++(0,-0.2) |- (mux3);
    \draw [->, Arrowy] (tanh) -- (mux3);
    \draw [->, Arrowy] (minus) -- (mux1);
  
    \draw [->,Arrowy] (mux3) -- (add1);



    \draw [->, Arrowy] (add1)++(0,0.28) -> (h2);
    \draw [ArrowC1] (b2) -|  (tanh);
    \draw [ArrowC1] (b1) -|  (sigma1);

\end{tikzpicture}
	  \caption{GRU cell.}
	  \label{fig:gru}
	 \end{figure}
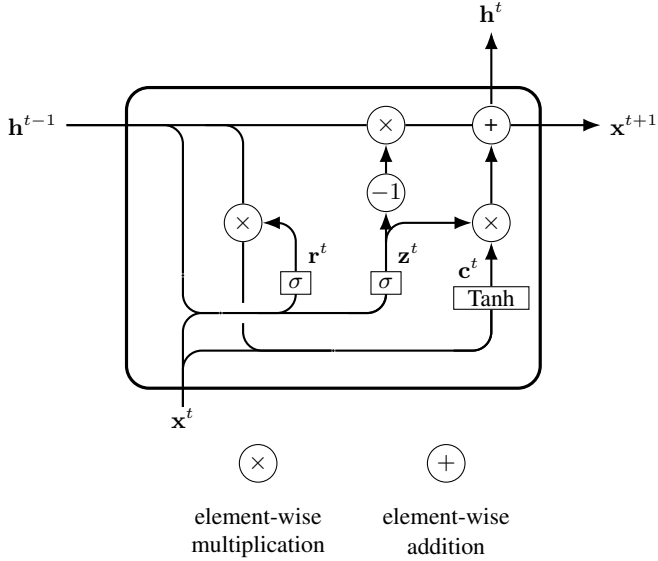

The proposed learning architecture employs a GRU cell~\cite{cho2014properties} as the basic building block. The GRU cell is composed by a hidden state $\mathbf{h}^t$, which allows the weights of the GRU to be shared across time, as well as by two  gates  $\mathbf{q}^t$ and $\mathbf{r}^t$, which modulate the flow of information inside the cell unit. Fig. \ref{fig:gru} depicts the architecture of a GRU. The gates are given by:
\begin{align}
\mathbf{q}^t &= \sigma(\mathbf{W}_q\mathbf{x}^t + \mathbf{V}_q\mathbf{h}^{t-1} + \mathbf{b}_q),\label{eq:gate1}\\
\mathbf{r}^t &= \sigma(\mathbf{W}_r\mathbf{x}^t + \mathbf{V}_r\mathbf{h}^{t-1} + \mathbf{b}_r),\label{eq:gate2}
\end{align}
where  $\{\mathbf{W}_q, \mathbf{V}_q,\mathbf{W}_r,\mathbf{V}_r\}\subset\mathbb{R}^{M\times M}$ are matrices whose entries are learnable weights, $\{\mathbf{b}_q,\mathbf{b}_r\}\subset\mathbb{R}^{M}$ are the bias parameters, and $\sigma(\cdot)$ is the sigmoid function. 

The update of the hidden state $\mathbf{h}^t$  is a linear combination of the previous hidden state and the candidate state $\mathbf{c}^t $:
\begin{align}
\mathbf{c}^t &= \sigma(\mathbf{W}_c\mathbf{x}^t + \mathbf{V}_c(\mathbf{h}^{t-1}\odot\mathbf{r}^t) + \mathbf{b}_c),\\
\mathbf{h}^t &= \mathbf{q}^t\odot\mathbf{c} + (1-\mathbf{q}^t)\odot\mathbf{h}^{t-1},
\end{align}	
with $\odot$ being the element-wise multiplication.
Similar to LSTM~\cite{graves2013generating}, the additive update of the hidden state can handle long-term dependencies by avoiding a quick vanishing of the errors in back-propagation, and by not overwriting important features. The GRU structure was chosen to compose the forecasting module of the proposed method because its performance is usually on pair with LSTM, but with a lower computational burden~\cite{Chung2014}; nonetheless, it could be replaced by an LSTM or any other type  of RNN. 
	 
	\subsection{Forecasting Module}
	
	The proposed learning model, named spectral graph GRU (SG-GRU), combines a standard  GRU cell applied to the vertex-domain GSs comprising $\mathbf{X}_{\mathcal{S}}^t$  with a GRU cell applied to the frequency-domain versions of the latter GSs comprising $\hat{\mathbf{X}}_{\mathcal{F}}^t$. The GRU acting on frequency-domain signals is named here spectral GRU (SGRU), and has the same structure as the standard GRU, except for the dimension of weight matrices and bias vectors, which are $K\times K$ and $K\time 1$, respectively. The dimension of the hidden state is therefore $K$.
	
	Assuming that the entire GS $\mathbf{x}$ is $(\mathcal{F},\epsilon)$-bandlimited, most of the information about it is expected to be stored in $\hat{\mathbf{x}}_{\mathcal{F}}$. Then, given an admissible operator $\boldsymbol{\Psi}_{\mathcal{S}}$, one has
	\begin{align}
	    \|\mathbf{x}-\boldsymbol{\Phi}_{\mathcal{S}}\boldsymbol{\Psi}_{\mathcal{S}}\mathbf{x}\|_2\leq \frac{\epsilon}{{\rm SV}_{\min}(\boldsymbol{\Psi}_{\mathcal{S}}\mathbf{U}_{:,\mathcal{F}})}.
	\end{align}
	The choice of $\mathcal{F}$ will be further discussed in the experiments described in Section~\ref{sec:results}.
	
	The SGRU module in the proposed learning framework is able to predict the (possibly time-varying) graph-frequency content of the network signals. This is key to fully grasping the underlying spatial information embedded in the graph frequency content. Besides, 
	it is worth pointing out that the proposed SGRU is slightly different from simply combining the spectral graph convolution (SGC) in~\cite{Bruna2014} with a GRU: in both cases, the input signal is previously transformed to the Fourier domain, but in the SGRU a standard GRU, composed by matrix-vector multiplications, is applied to the transformed signal, whereas in the latter case, the SGC computes the graph convolution, which is an element-wise vector multiplication. Thus, the SGRU is able to better capture the temporal relations among different spectral components.
	  
	\subsection{GS Interpolation}
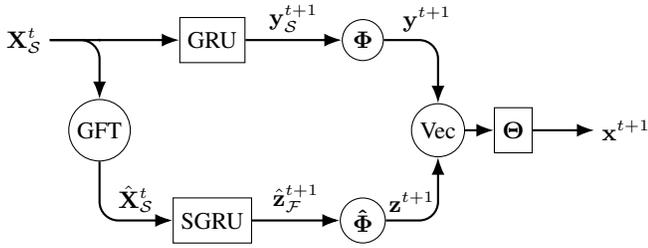
\begin{figure}
	  \footnotesize
		\centering
	  \begin{tikzpicture}[
    font=\small,
    >=LaTeX,
    cell/.style={
        rectangle, 
        rounded corners=4mm, 
        draw,
        very thick,
        },
    operator/.style={
        circle,
        draw,
        inner sep=2pt,
        minimum height =.5cm,
        },
    branch/.style={
        circle,
        fill,
        inner sep=0pt,
        },
    ct/.style={
        circle,
        draw,
        line width = .75pt,
        minimum width=1cm,
        inner sep=3pt,
        },
    gt/.style={
        rectangle,
        draw,
        minimum width=4mm,
        minimum height=6mm,
        inner sep=3pt
        },
    ArrowC1/.style={
        rounded corners=.25cm,
        thick,
        },
    Arrowy/.style={
        rounded corners=.5cm,
        thick,
        },
        cross/.style= {draw, fill=white, circle, node distance=1cm, minimum size=5pt}
    ]


	  \node (input) at (0.5,0) {$\mathbf{X}_{\mathcal{S}}^t$};
	  \node [branch] (branch) at (1.5,0)  {};
	  \node [gt] (gru) at (3,0)  {GRU};
	  \node [operator] (gft) at (1.5,-1.2)  {GFT};
	  \node [gt] (sgru) at (3,-2.4) {SGRU};
	  \node [operator](igft) at (5,0) {$\boldsymbol{\Phi}$};
	  \node [operator] (sigft) at (5,-2.4)  {$ \boldsymbol{ \hat{\Phi}}$};
	  \node [operator] (dot) at (6,-1.2)  {Vec};	 
	 
	  \node [gt] (system) at (7,-1.2)    {$\boldsymbol{ \Theta}$};

	  \node  (output)  at (8.5,-1.2) {$\mathbf{x}^{t+1}$};

	  \draw [->,ArrowC1] (input) -- node {} (gru);
	  \draw [->,ArrowC1] (input) -| node {} (gft);
	  \draw [->,ArrowC1] (gft) |- node[pos=0.75,above] {$\hat{\mathbf{X}}_{\mathcal{S}}^t$} (sgru);
	  
	  \draw [->,ArrowC1] (gru) -- node[above] {$\mathbf{y}_{\mathcal{S}}^{t+1}$} (igft);
	  \draw [->,ArrowC1] (sgru) -- node[above] {$\hat{\mathbf{z}}_{\mathcal{F}}^{t+1}$} (sigft);
	  \draw [->,ArrowC1] (igft) -| node[pos=0.4,above] {$\mathbf{y}^{t+1}$} (dot);
	  \draw [->,ArrowC1] (sigft) -| node[pos=0.25,above] {$\mathbf{z}^{t+1}$} (dot);	  
	  \draw [->,ArrowC1] (dot) -- node {}  (system);
	  \draw [->,ArrowC1] (system) -- node {}  (output);

\end{tikzpicture}
	  \caption{Proposed SG-GRU model. The input GS follows two routes in parallel: in the upper route, the GRU followed by  interpolation is applied to the GS; in the bottom route, the GS is transformed to frequency-domain before being processed by the SGRU module and thereafter being interpolated. The outputs of these two parallel processes are stacked into a single vector, represented by operation ``Vec'', and fed to an FC layer.}
	  \label{fig:model}
	    \end{figure}

	 The outputs from the GRU, $\mathbf{y}_{\mathcal{S}}^{t+1}$, and from the SGRU ,$\hat{\mathbf{z}}_{\mathcal{F}}^{t+1}$, are interpolated by $\boldsymbol{\Phi}_{\mathcal{S}}$ and $\hat{\boldsymbol{\Phi}}_{\mathcal{S}} = \boldsymbol{\Phi}_{\mathcal{S}}\mathbf{U}_{:,\mathcal{F}}$, respectively. The resulting $N$-dimensional vectors $\mathbf{y}^{t+1}$ and $\mathbf{z}^{t+1}$ are stacked in a single vector of size $2N$ which is processed by a fully connected (FC) layer to yield
	 \begin{align}
	 \tilde{\mathbf{x}}^{t+1} = \Theta(\mathbf{y}^{t+1} ,\mathbf{z}^{t+1}),\label{eq:output}
	 \end{align}	 
	  as illustrated in  Fig.~\ref{fig:model}.

	\subsection{Loss Function}
	The loss function employed is  the (empirical) mean square error (MSE). In a supervised scenario, the signal from all nodes are available for training, thus enabling the use of the entire GS $\mathbf{x}^{t+1}$ as label to compute the loss function. Given the batch size $T_{\rm b}$, the loss function for the supervised training $\mathcal{L}_{\rm s}$ is 
	\begin{align}\label{eq:loss-super}
	\mathcal{L}_{\rm s} = \frac{1}{T_{\rm b}N}\sum\limits_{t=1}^{T_{\rm b}}\|\tilde{\mathbf{x}}^{t+1}-\mathbf{x}^{t+1}\|_2^2.
	\end{align}
	
	 In the semi-supervised task, on the other hand, only the sampled ground-truth signal $\mathbf{x}^{t+1}_{\mathcal{S}}$ can be accessed. In order to achieve better predictions on the unknown nodes, we propose to interpolate the sampled ground-truth signal by $\boldsymbol{\Phi}_{\mathcal{S}}$ before computing the MSE, yielding
	\begin{align}\label{eq:loss-semi}
	\mathcal{L}_{\rm ss} =& \frac{1}{T_{\rm b}N}\sum\limits_{t=1}^{T_{\rm b}} \|\tilde{\mathbf{x}}^{t+1}-\mathcal{I}(\mathbf{x}^{t+1}_{\mathcal{S}})\|_2^2,
	\end{align}
	where 
	\begin{align}
	\displaystyle \left[\mathcal{I}(\mathbf{x_{\mathcal{S}}})\right]_n = \left\{\begin{array}{cc}
		x_n, & {\rm if \;} v_n \in \mathcal{S}\\
		\left[ \boldsymbol{\Phi}_{\mathcal{S}}\mathbf{x}^{t+1}_{\mathcal{S}}\right]_n, & {\rm otherwise.}
		\end{array}\right. \label{eq:interp_loss}
	\end{align} 
	
\subsection{Computational Complexity}
The SG-GRU consists of two GRU cells, refereed as GRU and SGRU, which compute $6$ matrix-vector multiplications each. The dimensions of the weight matrices in these recurrent modules applied on the vertex and frequency domains are  $M^2$ and $K^2$, respectively, where $K$ was set to $\frac{M}{3}$ in this paper (this choice will be further discussed in Section~\ref{sec:results}). The input of the SGRU is the sampled GS in the frequency domain, obtained by applying the truncated GFT, which is a $K\times M$ matrix. This transform can be pre-computed, avoiding the matrix vector multiplication during the loop recurrence. In this case the input of the network becomes a signal with dimension $M+K$. The output of the GRU and the SGRU are, thereafter, interpolated by $N\times M$ and $N\times K$ matrices, respectively, which are pre-computed before running the model. Finally, an FC layer is applied to the interpolated signals, costing $2N^2$ flops. Note that the truncated GFT, the interpolations, and the FC layers are out of the recurrence loop and do not increase the computational cost if a larger sequence length $\tau$ is used. Thus, the computational cost per iteration of the SG-GRU is
\begin{align}
    KM + 6\tau(M^2+K^2) + N(K+M) + 2N^2 \;\text{[flops]}.
\end{align}

\section{Applications}\label{sec:applications}

	The proposed learning architecture in  Fig.~\ref{fig:model} can handle both supervised and semi-supervised scenarios. 
	In the supervised case, measurements from the $N$ network nodes are available in the training step but not necessarily for testing. This supervised scenario covers many different applications; a case in point is a weather station network wherein the temperature sensors are working during a period of time, but then, suddenly, some of them are shut down due to malfunctioning or maintenance cost reduction. In the semi-supervised case, on the other hand, only part of the nodes appear in the training set and can, therefore, be used to compute gradients. Again, the semi-supervised scenario also covers many practical applications; for instance, when a sensor network is deployed with a limited number of nodes to reduce the related costs, but a finer spatial resolution is desirable, which can be obtained by a virtual denser sensor network.
	
	Considering these two basic scenarios, we can conceive four specific types of applications:
	
	\subsection{Supervised Application}\label{sub:supApp}
	
	Input GS is composed by $M \leq N$ nodes but labels of all $N$ nodes are used to compute the loss function in~\eqref{eq:loss-super}. As mentioned before, this learning model can be applied to situations in which  all the $N$ sensors are temporarily activated and, afterwards, $N-M$ sensors are turned off.
	
	\subsection{Semi-supervised Application}\label{sub:semiApp}
	
	Both input GS and labels are composed by $M<N$. Thus, only the $M$ in-sample are available to train the model using the loss function in~\eqref{eq:loss-semi}. In this application, it is desired to predict the state of a static network with $N$ nodes, considering that only $M<N$  sensors are deployed.
	
	\subsection{Noise-corrupted Application}\label{sub:noiseApp}
	
	Input GS is composed by all the $N$ nodes with signals corrupted by uncorrelated additive noise, and the labels are the entire ground-truth GS. This application allows working with the proposed learning model when the sensors' measurements are not accurate. In this case, only the denoising capacity of the proposed model is evaluated, hence no sampling is performed over the input data.
	
	\subsection{Missing-value Application}\label{sub:missApp}
	
	Input GS is composed by all the $N$ nodes but, at each time instant, a fraction of the $N$ values measured by the sensor network are randomly chosen to be replaced by {\tt NaN} (not a number). It is worth highlighting that this application is different from the (pure) supervised application in Section~\ref{sub:supApp}. In the supervised scenario, the set of known nodes, $\mathcal{S}$, is fixed across time, whereas the application of missing values considers different sets of known signal values at each time instant $t$. In other words, we have a supervised scenario with a time-dependent sampling set $\mathcal{S}^t$. The labels are the entire ground-truth GS. This setup evaluates the performance of the proposed SG-GRU when some of the sensors' measurements are missing, which could be due to transmission failures in a wireless network.

	\section{Numerical Experiments}\label{sec:results}

In this section,  we assess the performance of the proposed SG-GRU scheme in two real datasets. The simulation scenarios are instances of the four applications described in Section~\ref{sec:applications}. 

\subsection{Dataset Description}\label{sub:dataset}
	The proposed learning model was evaluated on two distinct multivariate time-series datasets: temperatures provided by the Global Surface Summary of the Day Dataset (GSOD),  which can be accessed at~\cite{gsod}, and the Seattle Inductive Loop Detector Dataset (SeattleLoop)~\cite{Cui2018}. 
	 
	\subsubsection{Global Surface Summary of the Day Dataset}
	The  GSOD dataset consists in daily temperature measurements in $^{\rm o}$C from $2007$ to $2013$, totalling $2,557$ snapshots, in $430$ weather stations distributed in the United States.\footnote{Weather stations in the Alaska and in Hawai were not considered.} The source provides more weather stations but only $430$ worked fully from $2007$ until $2013$. These stations are spatially represented by a $10$-nearest-neighbor graph with  nonzero edge weights given by~\cite{Sandryhaila2014}:
	\begin{align}
	A_{nm} \!=\! \frac{e^{-(d_{nm}^2 + h_{nm}^2)}}{\sqrt{\sum_{j \in \mathcal{N}_n}e^{-(d_{nj}^2 + h_{nj}^2)}}\sqrt{\sum_{j \in \mathcal{N}_m}e^{-(d_{mj}^2 + h_{mj}^2)}}},
	\end{align}
	in which $\mathcal{N}_n$ is the set of neighboring nodes connected to the node indexed by $n$, whereas $d_{nm}$ and $h_{nm}$ are, respectively, the geodesic distance  and the altitude difference between weather stations indexed by $n$ and $m$. The adjacency matrix is symmetric and the diagonal elements are set to zero.
	
	\subsubsection{Seattle Inductive Loop Detector Dataset} 
	The SeattleLoop dataset contains traffic-state data  collected from inductive loop detectors deployed on four connected freeways in the Greater Seattle area. The $323$ sensor stations measure the average speed, in miles$/$hour, during the entire year of $2015$ in a $5$-minute interval, providing $105,120$ timesteps. This dataset is thus much larger than GSOD.
	The graph adjacency matrix provided by the source~\cite{Cui2018} is binary and the GS snapshots are barely bandlimited with respect to the graph built on this adjacency matrix. To build a network model in which the SeattleLoop time series is $(\mathcal{F},\epsilon)$-bandlimited with a reasonably small $\epsilon$, the nonzero entries of the binary adjacency matrix were replaced by the radial-basis function 
	\begin{align}
	    A_{nm} = {\rm e}^{-\frac{\|\mathbf{x}_n-\mathbf{x}_m\|^2}{10}},
	\end{align}
	where $\mathbf{x}_n$ and $\mathbf{x}_m$ are time series, containing $1000$ time-steps, corresponding to nodes $v_n$ and $v_m$, respectively.

\subsection{Choice of Frequency Set $\mathcal{F}$}\label{sub:choiceF}
	The larger the set $\mathcal{F}$ the more information about the input signal is considered in the model. However, the interpolation using~\eqref{eq:recovery} is admissible only if $|\mathcal{F}|=K\leq M$~\cite{Chen2015}. Moreover, if $K$ increases, the smaller singular value of $\mathbf{U}_{\mathcal{S},\mathcal{F}}$  tends to decrease, leading to an unstable  interpolation. Since the GSs considered in this paper are approximately bandlimited, using $K$ close to $M$ accumulates error during the training of the network. Based on validation loss, $K$ was set to $\frac{M}{3}$.
	
	When all nodes are available for training, that is, in the applications described in Sections~\ref{sub:supApp},~\ref{sub:noiseApp}, and~\ref{sub:missApp},    $\mathcal{F}$ is chosen as the $K$ Laplacian eigenvalues corresponding to the dominant frequency components (the ones with highest energy) of signals measured at the first $100$ days. In the semi-supervised approach, on the other hand, the spectral content of the entire GS is unknown. Since the GSs considered in this paper are usually smooth, in the sense that most of their frequency content is supported on the indices associated with the smaller Laplacian eigenvalues, the set $\mathcal{F}$ was chosen as the $K$ smallest eigenvalues $\lambda_n$ in this scenario. The set $\mathcal{F}$ used in the application described in Section~\ref{sub:semiApp} is, therefore, slightly different from the set $\mathcal{F}$ used in the scenarios related to the applications of  Sections~\ref{sub:supApp},~\ref{sub:noiseApp}, and~\ref{sub:missApp}.

		\subsection{Competing Learning Techniques}\label{sub:benchmarks}	
	Recently many DL-based models were shown to outperform classical methods in the task of predicting ST data. Nonetheless, to the best of our knowledge, only~\cite{Khodayar2019} addresses the problem of predicting ST data by training a learning model with $M<N$ nodes, with the aim of reducing the training time duration.  Therefore, the performance of our proposed method is here compared with DL-based models from the literature that do not actually handle sampled input GSs. Thus, we adapted the DL-based models from the literature by combining them with an interpolation strategy, such as $k$-NN and the GSP-based interpolator $\boldsymbol{\Phi}_{\mathcal{S}}$. In this context, the interpolation can be performed either: (i) before running the forecasting technique, so that the input of the competing DL-based model will be the entire GS; or (ii) after running the forecasting technique, so that the input of the competing DL-based model will be a sampled GS, thus requiring fewer learnable  parameters.
	
	We use as benchmark some LSTM-based NNs, which were shown to perform well in the strict forecasting task (i.e., time-domain prediction) on the SeattleLoop dataset in comparison with other baseline methods, such as ARIMA and SVR~\cite{Cui2019}. In addition, we  also consider the ST graph convolution network (STGCN) proposed in~\cite{Yu2018} as benchmark. In summary, the competing techniques  (adapted to deal with sampled GSs) are:
	\begin{enumerate}[label=(\roman*)]
		\item  LSTM: simple LSTM cell;
		\item C1D-LSTM: a $1$D convolutional layer followed by an LSTM cell;
		\item SGC-LSTM: the SGC from~\cite{Bruna2014} followed by an LSTM;
		\item TGC-LSTM: a traffic graph convolution based on the adjacency matrix combined with LSTM~\cite{Cui2019};\footnote{Code from
		https:$//$github.com$/$zhiyongc$/$Graph$\_$Convolutional$\_$LSTM .}
		\item STGCN: a combination of the graph convolution from~\cite{Defferrard2016a} with a gated-temporal convolution~\cite{Yu2018}; Hyperparameters were set as in~\cite{Yu2018} since they lead to smaller MSE in the validation set (filter sizes were evaluated from the set $\{16,32,64\}$).\footnote{Code from https:$//$github.com$/$VeritasYin$/$Project$\_$Orion.} 
	\end{enumerate}
	 As mentioned before, the above competing techniques do not tackle joint forecasting and interpolation tasks. Thus, they were combined with an interpolation technique. The output of methods (i)-(iv) were interpolated by $\boldsymbol{\Phi}_{\mathcal{S}}$, whereas a $1$-hop neighborhood interpolation was applied before the method (v), that is, each unknown value $x^t_n$ was set as
	 \begin{align}
        [x^t_n]_{\rm unknown} = \frac{1}{|\mathcal{N}_n|}\sum_{m \in \mathcal{N}_n}x_m^t\,. \label{eq:1_hop_interp}
	 \end{align} 
	 Unlike LSTM-based methods, the interpolation in~\eqref{eq:1_hop_interp} provided better results when combined with STGCN to handle the sampled input GS. The TGC-LSTM was only applied to the SeattleLoop dataset since it uses a free-flow reachability matrix, being specifically designed for traffic networks.
	 
	 TABLE \ref{tab:models_comparison} summarizes the competing learning techniques along with the corresponding interpolation methods. ``GSP interpolation'' stands for interpolation by $\boldsymbol{\Phi}_{\mathcal{S}}$ and ``$1$-hop interpolation'' stands for interpolation by averaging records on the $1$-hop neighborhood. Also the order followed by each procedure is indicated: ``interpolation first'' stands for interpolating the data before running the learning model, whereas ``model first'' refers to running the model and then interpolating its output.
	 
	 \begin{table}[h]
	 	\centering
		\caption{Summary of competing techniques}
		{\def\arraystretch{2}\tabcolsep=10pt
	 	\begin{tabular}{|l|c|c|}
	 		\hline
	 		\bf Model & \bf Interpolation& \bf Order of procedures \\ \hline
	 		LSTM & GSP interpolation& model first \\\hline
	 		C1D-LSTM & GSP interpolation&model first \\\hline
	 		SGC-LSTM & GSP interpolation&model first\\\hline
	 		TGC-LSTM & GSP interpolation&model first\\\hline
	 		STGCN & $1$-hop interpolation&interpolation first\\\hline
	 		
	 	\end{tabular}}
 	\label{tab:models_comparison}
	 \end{table}

	\subsection{Figures of Merit}\label{sub:FoMs}
	The prediction performance was evaluated by the root mean square error (RMSE) and the mean absolute error (MAE):
	\begin{align}
	{\rm RMSE} = \frac{1}{T_{\rm t}} \sum_{t=1}^{T_{\rm t}} \left(\sqrt{\frac{1}{N}\sum_{n=1}^N (e_{n}^t)^2 }\right),
	\end{align}
	
	\begin{align}
	{\rm MAE} = \frac{1}{T_{\rm t}} \sum_{t=1}^{T_{\rm t}} \left(\frac{1}{N}\sum_{n=1}^N |e_{n}^t| \right),
		\end{align}
	where $T_{\rm t}$ is the number of test samples and $e_{n}^t$ is the prediction error of the $t^{\rm th}$ test sample and $n^{\rm th}$ node. In the noisy setup, the mean absolute percentage error (MAPE) was also evaluated:
	\begin{align}
	{\rm MAPE} = \frac{1}{T_{\rm t}} \sum_{t=1}^{T_{\rm t}} \left(\frac{1}{N}\sum_{n=1}^N \frac{|e_{n}^t|}{|x_n^t|}\right)\times 100\%\,.
	\end{align}
	The error metrics MAE and RMSE have the same units as the data of interest, but RMSE is more sensitive to large errors, whereas MAE tends to treat more uniformly the prediction errors. 
	 
	\subsection{Experimental Setup}\label{sub:setup}	
	In the applications described in Sections~\ref{sub:supApp} and~\ref{sub:semiApp}, $75\%$, $50\%$, and $25\%$ from the $N$ nodes in $\mathcal{V} $ were  selected to compose the set $\mathcal{S}$ using a greedy method of~\cite{winer1962statistical}, called E-optimal design, with the set $\mathcal{F}$ corresponding to the first $M$ smallest Laplacian eigenvalues. This choice of $\mathcal{F}$ relies on the smoothness of the underlying GS, that is, nodes near to each other are assigned with similar values. The same sampling sets were used for both supervised and semi-supervised training. All the experiments were conducted with a time window of length $\tau = 10$. The prediction length was $p=1$ and $p=3$ samples ahead for the GSOD dataset, that is, $1$ day and $3$ days, respectively,  and $p=1$ and $p=6$ samples to SeattleLoop, that is $5$ and $30$ minutes, respectively.
	
	The datasets were split into: $70\%$ for training, $20\%$ for validation, and $10\%$ for test. Batch size was set to $T_{\rm b} = 40$ and the learning rate was $10^{-4}$, with step decay rate of $0.5$ after every $10$ epochs. Training was stopped after $100$ epochs or $5$ non-improving validation loss epochs. The input of the model was normalized by the maximum value in the training set. The model was trained by the RMSprop~\cite{hinton2012neural} with PyTorch default parameters~\cite{rmsprop}. The network was implemented in PyTorch 1.4.0 and  experiments were conducted on a single NVIDIA GeForce GTX 1080.

\subsection{Results: Supervised Application}\label{sub:resSup}
	TABLE~\ref{tab:usa_sup} and TABLE~\ref{tab:seattle_sup} show the MAE and RMSE in the supervised application. The proposed method outperformed all competitors in virtually all scenarios. When the sample size decreases, the performance gap increases compared to the benchmarks. On the GSOD dataset, the SG-GRU performed much better than the other strategies. We can see that, as  the temperature GS is approximately ($\mathcal{F}$,$\epsilon$)-bandlimited with small $\epsilon$,  the SG-GRU successfully captures spatial correlations by predicting the GSs' frequency content. 
	\begin{table}[ht]		
\centering
\caption{MAE and RMSE of supervised prediction applied to the  GSOD dataset}
\scriptsize
\begin{tabular}{p{0.015\textwidth}|lcccccc}
	\toprule
	\multicolumn{1}{c}{}&\multicolumn{1}{l|}{}&\multicolumn{2}{|c|}{$M=0.75 N$} &\multicolumn{2}{|c|}{$M=0.50 N$} &\multicolumn{2}{|c}{$M=0.25 N$} \\ \midrule
	&\textbf{Methods}& \textbf{MAE} & \textbf{RMSE} & \textbf{MAE} & \textbf{RMSE} & \textbf{MAE} & \textbf{RMSE}\\
	\midrule
	 \multirow{5}{*}{$p$=1}&\textbf{SG-GRU} & $\boldsymbol{1.66}$ & $\boldsymbol{2.21}$ & $\boldsymbol{1.73}$ & $\boldsymbol{2.28}$ & $\boldsymbol{1.74}$ & $\boldsymbol{2.31}$\\
	&\textbf{LSTM} & 2.37 & 3.11 & 2.35 & 3.09 & 2.52 & 3.31\\
	&\textbf{C1D-LSTM} & 2.32 & 3.02 & 2.40& 3.15 & 2.66& 3.49\\
	&\textbf{SGC-LSTM} & 3.15& 4.15 & 3.20 & 4.25 & 3.23 & 4.28\\
	&\textbf{STGCN} & 2.20& 2.98 & 2.44 & 3.29 & 2.40 & 3.22\\
	\bottomrule
\end{tabular}
	\label{tab:usa_sup}
\end{table}
	\begin{table}[ht]		
	\centering
	\caption{MAE and RMSE of supervised prediction applied to the SeattleLoop dataset}
	\scriptsize
	\begin{tabular}{p{0.015\textwidth}|lcccccc}
		\toprule
		\multicolumn{1}{c}{}&\multicolumn{1}{l|}{}&\multicolumn{2}{|c|}{$M=0.75 N$} &\multicolumn{2}{|c|}{$M=0.50 N$} &\multicolumn{2}{|c}{$M=0.25 N$} \\ \midrule
		&\textbf{Methods}& \textbf{MAE} & \textbf{RMSE} & \textbf{MAE} & \textbf{RMSE} & \textbf{MAE} & \textbf{RMSE}\\
		\midrule
	    \multirow{6}{*}{$p$=1}&\textbf{SGGRU} &$\boldsymbol{2.79}$ & $\boldsymbol{4.16}$ & $\boldsymbol{3.02}$ & $\boldsymbol{4.59}$ & $\boldsymbol{3.38}$ & $\boldsymbol{5.40}$\\
		&\textbf{LSTM} & 3.15 & 4.79 & 3.64 & 5.59 & 4.45 & 7.03\\
		&\textbf{C1D-LSTM} & 3.25 & 4.95 & 3.70 & 5.70 & 4.49 & 7.08\\
		&\textbf{SGC-LSTM} & 3.59 & 5.57 & 3.97 & 6.14 & 4.60 & 7.26\\
		&\textbf{TGC-LSTM} & 3.03 & 4.59 & 3.54 & 5.45 & 4.40 & 6.98\\
		 &\textbf{STGCN}   & $\boldsymbol{2.79}$ & 4.32 & 3.11 & 4.82 & 3.65 & 6.10\\
		\bottomrule
	\end{tabular}
	\label{tab:seattle_sup}
\end{table}

\subsection{Results: Semi-supervised Application}\label{sub:resSemi}

\begin{figure}[!t]
	\subfloat[US weather stations from GSOD dataset. \label{fig:usa1a}]{\centerline{\includegraphics[width=\columnwidth]{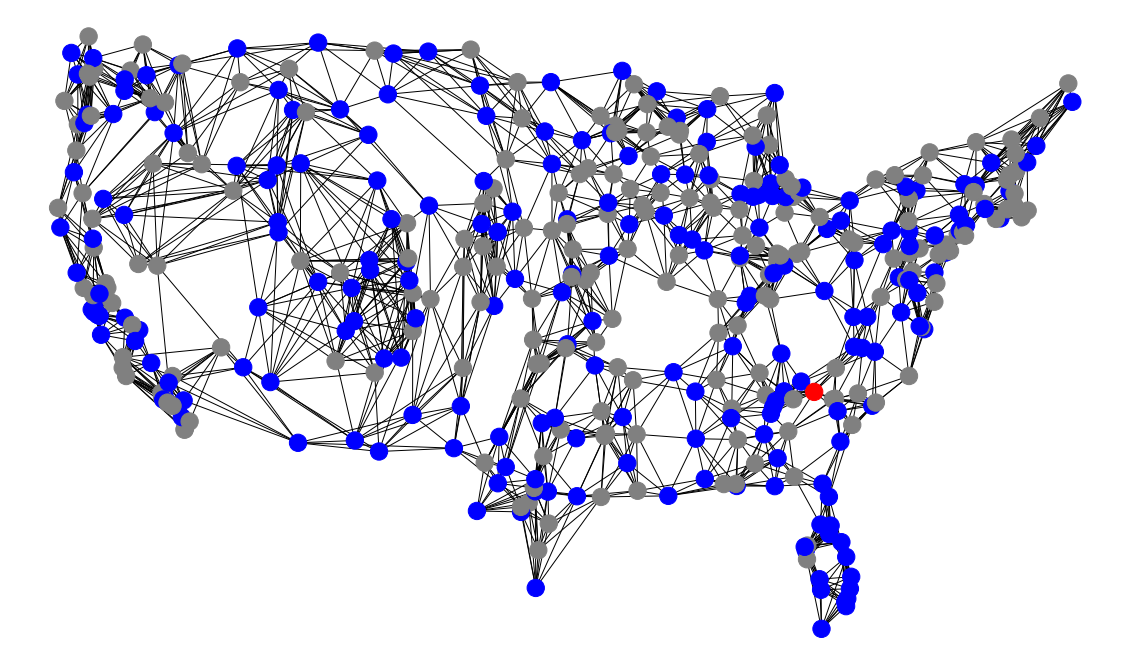}}}

	\subfloat[Predicted temperature on a single sensor. \label{fig:usa1b}]{\centerline{\includegraphics[width=\columnwidth]{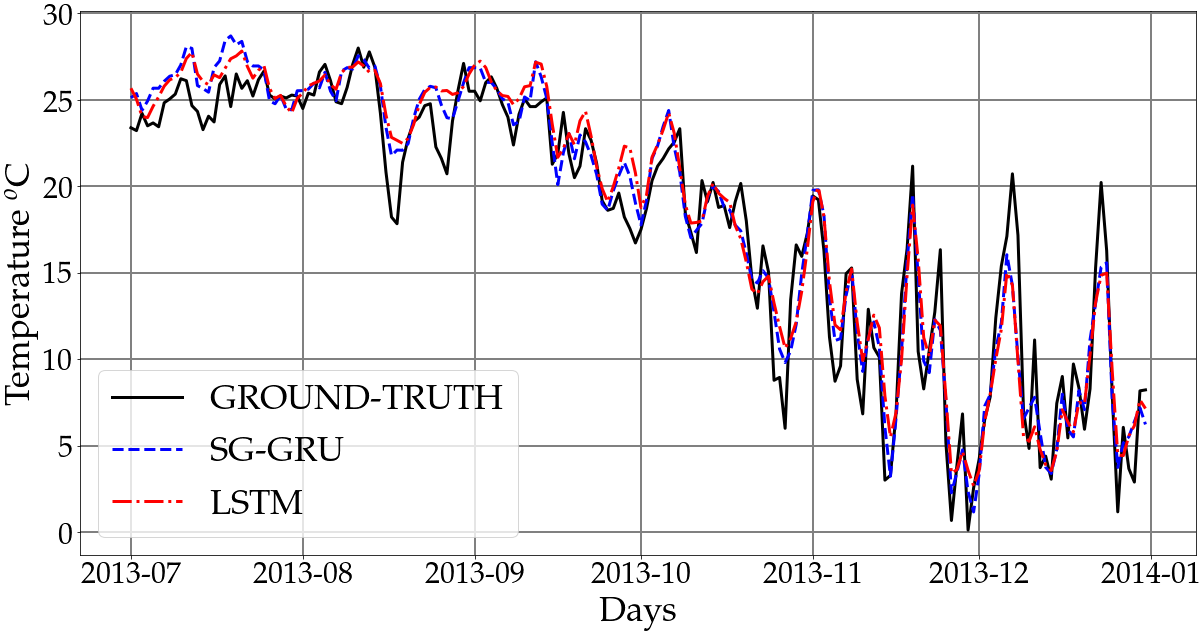}}}

	\caption{(a) Graph of sensors in the GSOD dataset. The known ($50\%$) and unknown ($50\%$) nodes are colored by blue and gray, respectively. The red node, which does belong to $\mathcal{S}$, indicates the weather station whose temperature predictions are shown in (b); and (b) output of the SG-GRU and the LSTM over the ground-truth temperature in the $2^{\rm nd}$ semester of 2013 measured by the node highlighted in red in (a).}
	\label{fig:temp}
\end{figure}

\begin{figure}[t]
	\subfloat[SG-GRU and TGC-LSTM predictions. \label{fig2:a}]{\centerline{\includegraphics[width=\columnwidth]{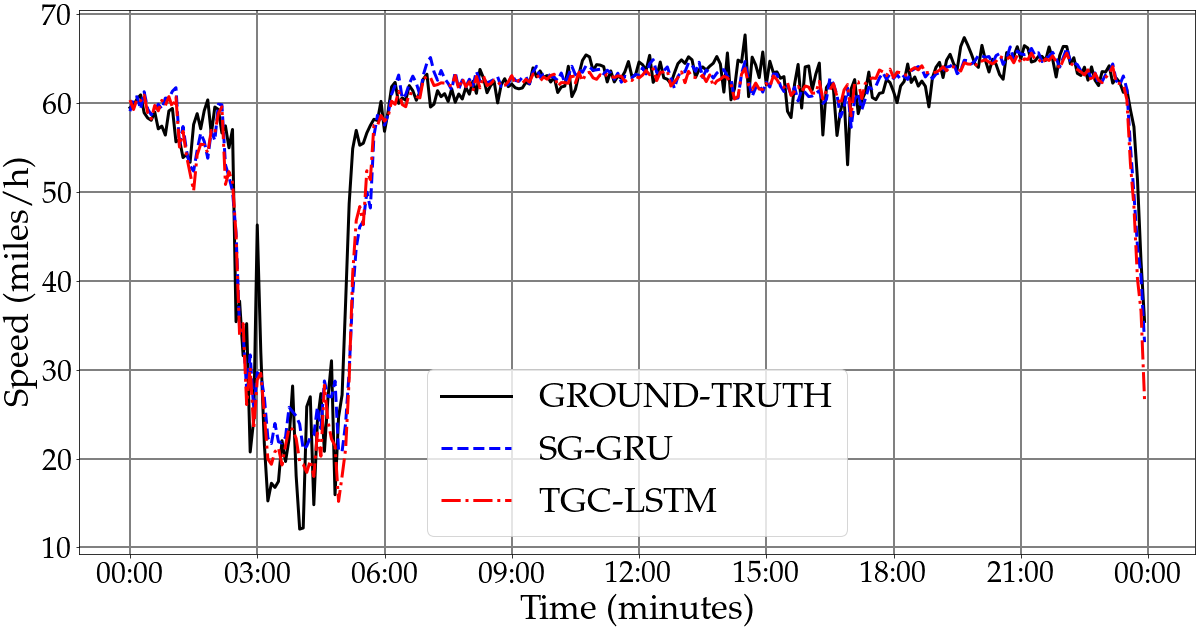}}}
	
		\subfloat[SG-GRU and STGCN predictions. \label{fig2:b}]{\centerline{\includegraphics[width=\columnwidth]{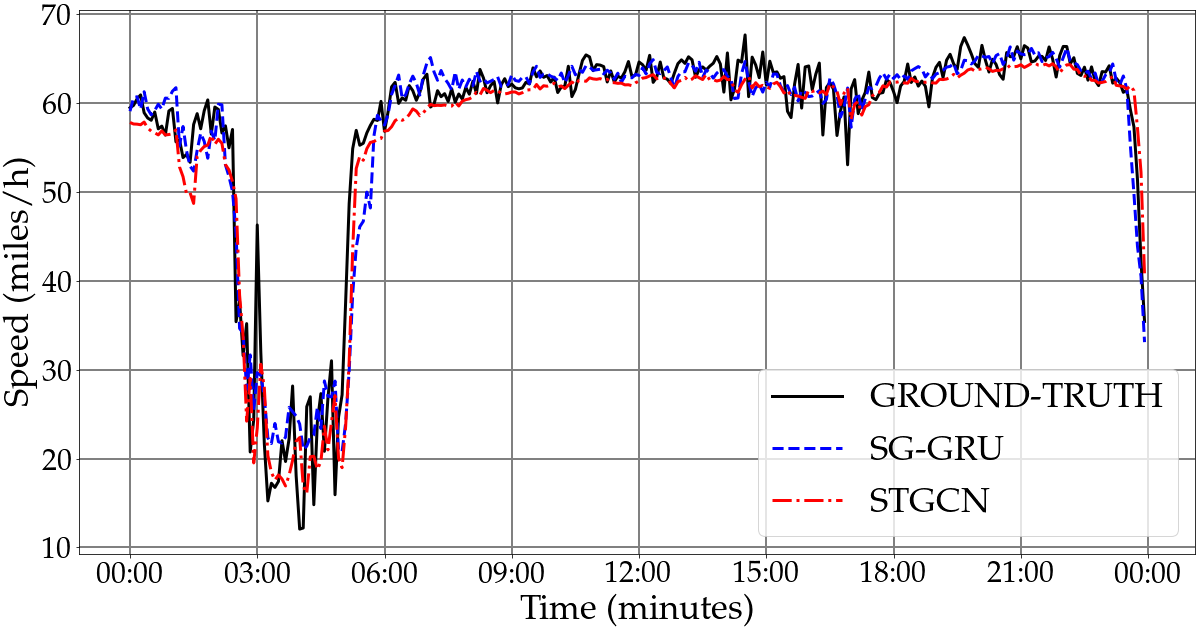}}}
	\caption{Predicted signal of the sensor i005es16920 using a subset  with $50\%$ of the nodes for the SG-GRU, TGC-LSTM, and STGCN. The evaluated sensor was absent in the sampling set $\mathcal{S}$.}
	\label{fig:seattle2}
\end{figure}

The loss function in~\eqref{eq:loss-semi} was used for training the SG-GRU and the LSTM-based methods. For the STGCN, the interpolation of the target GS in ~\eqref{eq:interp_loss} was replaced by the $1$-hop interpolation.
TABLE~\ref{tab:usa_semis} and TABLE~\ref{tab:seattle_semis} show the result of the SG-GRU and the competing approaches on the SeattleLoop and GSOD datasets, respectively. Fig.~\ref{fig:usa1b} plots the outputs of the SG-GRU and LSTM methods, in the second semester of 2013 over the ground-truth signal, for a weather station out of the sampling set, highlighted in Fig. \ref{fig:usa1a}, considering a situation with $50\%$ of known nodes. 
The SG-GRU outperformed the competing methods in the GSOD dataset. Since temperature GSs are highly smooth in the graph domain, the GSP interpolation, which is based on the assumption that the GS is bandlimited, provides good reconstruction. The energy of SeattleLoop dataset, on the other hand,  is not as concentrated as the GSOD dataset, leading to a larger reconstruction error. Even with this limitation on the prior smoothness assumption, the SG-GRU outperformed the STGCN combined with $1$-hop interpolation and the TGC-LSTM combined with GSP interpolation when the sampling  set size  is $25\%$ or  $50\%$ of the total number of nodes. It is worth mentioning that the STGCN and  the TGC-LSTM are  learning models designed specifically for traffic forecasting. When the horizon of prediction is $30$ minutes, then the SG-GRU achieved the smallest errors among all other methods. This could be due to simultaneous ST features extraction by the SGRU module. Fig.~\ref{fig:seattle2} depicts the predicted speed by SG-GRU, STGCN, and TGC-LSTM for an unknown sensor with $p=1$, $M=0.50N$, and during the day $11/24/2015$. As can be seen, the SG-GRU was able to better fit many points in the speed curve. It is worth mentioning that, despite the STGCN having poorly fitted the curve in Fig.~\ref{fig2:b}, it actually achieved higher accuracy on the known samples.

\begin{table}
	\centering	
	\caption{MAE and RMSE of semi-supervised prediction applied to the GSOD dataset}
	\scriptsize
\begin{tabular}{p{0.015\textwidth}|lcccccc}
	\toprule
	\multicolumn{1}{c}{}&\multicolumn{1}{l|}{}&\multicolumn{2}{|c|}{$M=0.75 N$} &\multicolumn{2}{|c|}{$M=0.50 N$} &\multicolumn{2}{|c}{$M=0.25 N$} \\ \midrule
	&\textbf{Methods}& \textbf{MAE} & \textbf{RMSE} & \textbf{MAE} & \textbf{RMSE} & \textbf{MAE} & \textbf{RMSE}\\
	\midrule\multirow{5}{*}{$p$=1}&\textbf{SG-GRU} & $\boldsymbol{1.77}$ & $\boldsymbol{2.38}$ & $\boldsymbol{1.88}$ & $\boldsymbol{2.53}$ & $\boldsymbol{2.06}$ & $\boldsymbol{2.76}$\\
	&\textbf{LSTM} & 2.35& 3.03 & 2.41 & 3,16 & 2.72 & 3.54\\
	&\textbf{C1D-LSTM} &1.83& 2.44 & 2.00& 2.65 & 2.24 & 2.97\\
	&\textbf{SGC-LSTM}  & 2.75 & 3.66 & 2.84 & 3.76 & 3.01 & 3.97\\
	&\textbf{STGCN} & 2.34& 3.2 & 3.75 & 5.02 & 6.92 & 8.65\\\midrule
	\multirow{5}{*}{$p$=3}&
	\textbf{SG-GRU} & $\boldsymbol{2.84}$& $\boldsymbol{3.76}$ &$\boldsymbol{  2.90}$& $\boldsymbol{3.85}$ & $\boldsymbol{2.99}$& $\boldsymbol{3.94}$\\
	&\textbf{LSTM} &  2.88 & 3.83 & 2.95 & 3.92 & 3.04 & 4.03\\
	&\textbf{C1D-LSTM} & 2.88 & 3.84 & 2.96 & 3.92 & 3.05 & 4.03\\
	&\textbf{SGC-LSTM} & 3.12 & 4.15 & 3.16 & 4.20 & 3.28 & 4.36\\
	&\textbf{STGCN} & 3.33 & 4.40 & 4.28 & 5.53 & 6.95& 8.48\\
	\bottomrule
\end{tabular}
	\label{tab:usa_semis}
\end{table}
	
\begin{table}	
	\centering
	\caption{MAE and RMSE of semi-supervised approaches applied to the SeattleLoop dataset}
	\scriptsize
	\begin{tabular}{p{0.015\textwidth}|lcccccc}
		\toprule
		\multicolumn{1}{c}{}&\multicolumn{1}{l|}{}&\multicolumn{2}{c|}{$M=0.75 N$} &\multicolumn{2}{|c|}{$M=0.50 N$} &\multicolumn{2}{|c}{$M=0.25 N$} \\ \midrule
		&\textbf{Methods}& \textbf{MAE} & \textbf{RMSE} & \textbf{MAE} & \textbf{RMSE} & \textbf{MAE} & \textbf{RMSE}\\\midrule
		\multirow{6}{*}{$p$=1}&\textbf{SG-GRU} & 2.98 & 4.60 & $\boldsymbol{3.53}$ & $\boldsymbol{5.55}$ & $\boldsymbol{4.50}$ & $\boldsymbol{7.28}$\\
		&\textbf{LSTM} & 3.06 & 4.73 & 3.61& 5.66 & 4.56& 7.34\\
		&\textbf{C1D-LSTM} & 3.09 & 4.77 & 3.67 & 5.74 & 4.61& 7.4\\
		&\textbf{SGC-LSTM} & 3.46 & 5.38 & 3.86 & 5.99 & 4.65 & 7.44\\
		&\textbf{TGC-LSTM} &  3.01 & 4.61 & 3.64  & $\boldsymbol{5.55}$&  4.82&7.75\\
		&\textbf{STGCN}  & $\boldsymbol{2.88}$ & $\boldsymbol{4.65}$ & 3.72 & 6.46 & 5.67 & 10.3\\\midrule
		\multirow{6}{*}{$p$=6}&\textbf{SG-GRU} & $\boldsymbol{3.87}$ & $\boldsymbol{6.18}$ & $\boldsymbol{4.18}$ & $\boldsymbol{6.61}$ & $\boldsymbol{4.88}$ & $\boldsymbol{7.77}$\\
		&\textbf{LSTM} &  3.96 & 6.34 & 4.31 & 6.81 & 4.98 & 7.94\\
		&\textbf{C1D-LSTM} & 3.96 & 6.37 & 4.3 & 6.83 & 5.02 & 7.97\\
		&\textbf{SGC-LSTM} & 4.12& 6.63 & 4.44 & 6.98 & 5.03& 7.98\\
		&\textbf{TGC-LSTM} & 4.91 & 7.89 & 5.17 & 8.23 & 8.29& 12.5\\
		&\textbf{STGCN} & 4.54 & 6.82 & 4.56 & 7.90 & 6.11 & 10.8\\
		\bottomrule
	\end{tabular}
	\label{tab:seattle_semis}
\end{table}

\subsection{Results: Noise-corrupted Application}\label{sub:resNoise}
	In many real situations, sensors' measurements can be contaminated with noise, which may worsen forecasting accuracy. Therefore, to deal with these situations, it is important to develop robust algorithms. 	
	Consider a GS $\mathbf{x}$ with standard deviation $\sigma_x$ and a measurement Gaussian noise, uncorrelated across both time and graph-domain, $\boldsymbol{\eta}$ with standard deviation $\sigma_{\eta}$. The noisy GS is $\tilde{\mathbf{x} }= \mathbf{x} + \boldsymbol{\eta}$ if the whole network is measured or $\tilde{\mathbf{x}}_{\mathcal{S} }=  \boldsymbol{\Psi}_{\mathcal{S} }\mathbf{x} + \boldsymbol{\eta}$, if only the subset $\mathcal{S}$ is measured. 

	To evaluate the robustness of the proposed learning scheme, both  SeattleLoop and GSOD datasets were corrupted by additive Gaussian noise with zero mean and standard deviation (std) $\sigma_{\eta} = 0.5\sigma_x$ and $\sigma_{\eta} = 0.1\sigma_x$, where $\sigma_x$ is the std of the entire dataset: $10^{\rm o}{\rm C}$ for GSOD dataset and $12.74$ miles$/$h for SeattleLoop dataset. In this experiment, nodes were not sampled and only the capability of handling noisy input was evaluated.  TABLE~\ref{tab:usa_noise} and TABLE~\ref{tab:noise_seattle} show MAE, RMSE, and MAPE\footnote{Temperatures in the GSOD dataset were converted to Fahrenheit before computing MAPE to avoid division by zero.} of the forecasting models respectively evaluated on $100$ and $30$ simulations of each of these noisy scenarios. 
	
	In the GSOD dataset, the proposed model achieved reasonable error levels in the presence of noisy measurements: for instance, MAE and RMSE increased $9\%$ and $7\%$ in comparison with the supervised situation with $M = 0.75N$ when the additive noise has std $\sigma_{\eta} = 0.1\sigma_{x}$. Many GS denoising approaches are based on attenuating high frequencies of the GS~\cite{Shuman2011,Tremblay2016a}. The SGRU module of the proposed model promotes the smoothness of the predicted GS similarly: it runs a predictive algorithm over a restricted subset of the graph frequency content,  $\mathcal{F}$, and thereafter computes the inverse GFT considering only this restricted subset,     
	
	In the SeattleLoop dataset, the MAE and RMSE evaluated on the proposed model increased $4\%$ and $2\%$, respectively, in comparison with the supervised situation with $M = 0.75N$ when the additive noise has std $\sigma_{\eta} = 0.1\sigma_{x}$, This is a highly acceptable result, even though the STGCN achieved lower errors. 
	\begin{table}[!ht]
		\centering		
		\caption{MAE, RMSE and MAPE ($\%$) of  forecasting  applied to the GSOD  with noise corruption}
		\scriptsize
\begin{tabular}{p{0.015\textwidth}|lp{0.035\textwidth}p{0.035\textwidth}p{0.035\textwidth}p{0.035\textwidth}p{0.035\textwidth}p{0.035\textwidth}}
	\toprule
	\multicolumn{1}{c}{}&\multicolumn{1}{l|}{}&\multicolumn{3}{|c|}{$\sigma_{\eta}=0.1 \sigma_{x}$} &\multicolumn{3}{|c}{$\sigma_{\eta}= 0.5\sigma_{x}$} \\ \midrule 
	&\textbf{Methods}& \textbf{MAE} & \textbf{RMSE} & \textbf{MAPE} & \textbf{MAE} & \textbf{RMSE} & \textbf{MAPE}\\\midrule
	\multirow{6}{*}{$p$=1}&\textbf{SG-GRU} & 
	$\boldsymbol{1.81}$& $\boldsymbol{2.36}$ & $\boldsymbol{7.70}$ & $\boldsymbol{2.01}$& $\boldsymbol{2.61}$ & $\boldsymbol{8.52}$\\
	&\textbf{LSTM} & 1.98& 2.59 & 8.55 & 2.11 & 2.75 & 9.70\\
	&\textbf{C1D-LSTM} & 1.90 & 2.49 & 8.07 & 2.03 & 2.65 & 8.65\\
	&\textbf{SGC-LSTM} & 2.94& 3.89 & 13.9 & 2.95 & 3.91 & 13.9\\
	&\textbf{STGCN}&  2.19 & 2.94 & 10.3 & 2.66 & 3.48 & 12.3 \\\midrule

	\multirow{6}{*}{$p$=3}&\textbf{SG-GRU} &  $\boldsymbol{2.85}$ & $\boldsymbol{3.79}$ & $\boldsymbol{13.41}$ & $\boldsymbol{2.89}$ & $\boldsymbol{3.83}$ & $\boldsymbol{13.5}$\\
	&\textbf{LSTM}& 2.88 &  3.83& 13.6 & 2.93 & 3.88 & 13.8  \\
	&\textbf{C1D-LSTM} &  2.86 & 3.8 & 13.4 & 2.91 & 3.85 & 13.6 \\
	&\textbf{SGC-LSTM}& 3.18 & 4.21 & 15.2 & 3.16 & 4.2 & 15.2 \\
	&\textbf{STGCN} & 3.17 & 4.23 & 15.6 &  3.21 & 4.25 & 15.7 \\
	\bottomrule
\end{tabular}

\label{tab:usa_noise}
	\end{table}
	\begin{table}[!ht]
	\centering
	\caption{MAE, RMSE and MAPE ($\%$) of forecasting applied to the SeattleLoop with noise corruption }
	\scriptsize
\begin{tabular}{p{0.015\textwidth}|lp{0.03\textwidth}p{0.035\textwidth}p{0.035\textwidth}p{0.035\textwidth}p{0.035\textwidth}p{0.035\textwidth}}
	\toprule 
	\multicolumn{1}{c}{}&\multicolumn{1}{l|}{}&\multicolumn{3}{|c|}{$\sigma_{\eta}=0.1 \sigma_{x}$} &\multicolumn{3}{|c}{$\sigma_{\eta}=0.5 \sigma_{x}$} \\ \midrule 
	&\textbf{Methods} & \textbf{MAE} & \textbf{RMSE} & \textbf{MAPE} & \textbf{MAE} & \textbf{RMSE} & \textbf{MAPE}\\\midrule

	\multirow{6}{*}{$p$=1}&\textbf{SG-GRU} & 2.91 & 4.27 & 13.7 & 3.13 & 4.66 & 16.1\\
	&\textbf{LSTM} & 3.21 & 4.85 & 18.2 & 3.45 & 5.20 & 19.4\\
	&\textbf{C1D-LSTM} & 3.30 & 5.05 & 19.4 & 3.48 & 5.30 & 20.5\\
	&\textbf{SGC-LSTM} & 3.96 & 6.28 & 28.3 & 4.07 & 6.44 & 29.9\\
	&\textbf{TGC-LSTM} & 2.88 & 4.24 & 13.0 & 3.19 & 4.74 & 14.2\\
	&\textbf{STGCN}  & $\boldsymbol{2.63}$ & $\boldsymbol{3.85}$ &$\boldsymbol{11.3}$  & $\boldsymbol{3.08}$ & $\boldsymbol{4.39}$ & $\boldsymbol{12.9}$\\\midrule
	\multirow{6}{*}{$p$=6}&\textbf{SG-GRU}  & $\boldsymbol{3.74}$ & $\boldsymbol{6.02}$ & 26.3 & $\boldsymbol{3.84}$ & $\boldsymbol{6.19}$ & 26.1\\
	&\textbf{LSTM} & 3.99 & 6.35 & 28.6 & 4.07 & 6.47 & 27.1\\
	&\textbf{C1D-LSTM} & 3.99 & 6.39 & 28.6 & 4.07 & 6.49 & 28.1\\
	&\textbf{SGC-LSTM} & 4.56 & 7.27 & 34.6 & 4.58 & 7.29 & 34.7\\
	&\textbf{TGC-LSTM} & 3.79 & 6.09 & $\boldsymbol{26.1}$ & 3.92 & 6.28 & $\boldsymbol{25.3}$\\
	&\textbf{STGCN}   &3.77 & 6.18 & 26.4 & 4.00 & 6.33 & 26.0\\
	\bottomrule
\end{tabular}
\label{tab:noise_seattle}
\end{table}

\subsection{Results: Missing-value Application}\label{sub:resMiss}

Another common problem in real time-series datasets are missing values, which could occur due to sensor's malfunctioning  or failure in transmission. To evaluate the performance of the SG-GRU in this situation, $10\%$ of both SeattleLoop and GSOD datasets were randomly set to {\tt NaN}. Before applying the forecasting methods, each {\tt NaN} value, $x^t_n$, was  interpolated by the $1$-hop interpolation in \eqref{eq:1_hop_interp}. TABLE~\ref{tab:nan_usa} and TABLE~\ref{tab:nan_seattle} show the numerical results of this scenario considering two forecasting horizons on the GSOD and SeattleLoop datasets, respectively. The forecasting accuracy decreases when there are missing values, as expected. For instance, in the GSOD dataset, MAE and RMSE increased $6\%$ and $4\%$ in comparison with the supervised situation with $M = 0.75N$. In the SeattleLoop dataset, MAE increases about $10\%$ whereas the RMSE decreases about $8\%$. The GFT in the proposed model (and also in combination with the LSTM-based models) tends to smooth the output signal, reducing large deviations and consequently the RMSE. Nonetheless it can slightly increase the forecasting error across many nodes, leading to the increase in MAE.  
\begin{table}[!ht]
	
	\centering
	\caption{MAE, RMSE and MAPE ($\%$)  of forecasting applied to the GSOD dataset with  $10\%$ of missing values}
	\scriptsize
\begin{tabular}{p{0.015\textwidth}|lp{0.03\textwidth}p{0.035\textwidth}p{0.035\textwidth}p{0.035\textwidth}p{0.035\textwidth}p{0.035\textwidth}}
	\toprule
	\multicolumn{1}{c}{}&\multicolumn{1}{l|}{}&\multicolumn{3}{|c|}{$p=1$} &\multicolumn{3}{|c}{$p=3$} \\ \midrule 
	&\textbf{Methods}& \textbf{MAE} & \textbf{RMSE} & \textbf{MAPE} & \textbf{MAE} & \textbf{RMSE} & \textbf{MAPE}\\\midrule
	&\textbf{SG-GRU} & 
	$\boldsymbol{1.75}$& $\boldsymbol{2.3}$ & $\boldsymbol{7.53}$  &$\boldsymbol{2.87}$ & $\boldsymbol{3.77}$ &$\boldsymbol{13.1}$ \\
	&\textbf{LSTM}& 2.56 & 3.41 & 12.3&2.94 & 3.93 & 14.1\\	
	&\textbf{C1D-LSTM} & 2.53 & 3.36 & 11.9& 2.92 & 3.9 & 13.9\\	
	&\textbf{SGC-LSTM} & 3.51 & 4.81 & 20.2 & 3.22& 4.34 & 16.8\\
	&\textbf{STGCN}&  2.10 & 2.87 & 9.97&  3.22 & 4.31 & 15.2\\
	
	\bottomrule
\end{tabular}
	\label{tab:nan_usa}
\end{table}
\begin{table}[!ht]
	
	\centering
	\caption{MAE, RMSE and MAPE ($\%$)  of forecasting applied to the  SeattleLoop dataset with  $10\%$ of missing values}
	\scriptsize
\begin{tabular}{p{0.015\textwidth}|lp{0.03\textwidth}p{0.035\textwidth}p{0.035\textwidth}p{0.035\textwidth}p{0.035\textwidth}p{0.035\textwidth}}
	\toprule
	\multicolumn{1}{c}{}&\multicolumn{1}{l|}{}&\multicolumn{3}{|c|}{$p=1$} &\multicolumn{3}{|c}{$p=6$} \\ \midrule 
	&\textbf{Methods}& \textbf{MAE} & \textbf{RMSE} & \textbf{MAPE} & \textbf{MAE} & \textbf{RMSE} & \textbf{MAPE}\\\midrule
		 &\textbf{SG-GRU}  &3.10 & $\boldsymbol{3.85}$ & 4.60& $\boldsymbol{6.17}$ & 14.9&$\boldsymbol{23.7}$\\
		&\textbf{LSTM} & 3.37&4.07 & 5.08&6.49 & 19.6&27.3\\
		&\textbf{C1D-LSTM} & 3.44&4.06 & 5.23&6.49 & 19.6&27.3\\
		&\textbf{SGC-LSTM} & 4.01&4.58 & 6.34&7.31 & 16.3&35.1\\
		&\textbf{TGC-LSTM} & 3.15&3.91 & 4.70&6.26 & 13.5&24.8\\
		&\textbf{STGCN}    & $\boldsymbol{2.60}$ &3.91  &$\boldsymbol{3.95}$ &6.26 & $\boldsymbol{11.8}$ &24.9 \\
		\bottomrule
	\end{tabular}
	\label{tab:nan_seattle}
\end{table}
\subsection{Computational Cost and Efficiency}\label{sub:cost}

In the SeattleLoop Dataset, the epoch duration of  SG-GRU was, on average, $8.5$~s, whereas the more complex approaches, TGC-LSTM and STGCN, took around $40$~s and $84$~s per epoch, respectively. In the GSOD dataset, which is much shorter than the SeattleLoop, the average epoch duration of SG-GRU, LSTM, and STGCN were $0.20$~s, $0.25$~s, and $2.5$~s, respectively. TABLE~\ref{tab:time} shows the average training time, including pre-processing and data preparation, as well as  test phases for the 3 semi-supervised scenarios applied on the SeattleLoop  and GSOD datasets, with $p=1$. The SG-GRU required more epochs to converge than STGCN, but it still trains faster than the STGCN and also than the other competing approaches. 
\begin{table}[ht]
\caption{Average computational time  in seconds}
	 \label{tab:time}
    \centering
    
	 	{\def\arraystretch{1.5}\tabcolsep=8pt
	 	\begin{tabular}{lcccccc}
		\toprule
        &\multicolumn{2}{c}{SeattleLoop}&\multicolumn{2}{c}{GSOD}\\\midrule
		\textbf{Methods}& \textbf{Training} & \textbf{Test}& \textbf{Training} & \textbf{Test} \\
		\hline
		\textbf{SG-GRU}&414.68&4.89&11.18&0.01\\
		\textbf{LSTM}&1134.0&5.63&30.74&0.01\\
		\textbf{C1D-LSTM}&1319.0&5.90&36.90&0.03\\
		\textbf{SGC-LSTM}&2770.3&6.10&39.89&0.04\\
        \textbf{TGC-LSTM}&1027.1&5.48&-&-\\
		\textbf{STGCN}&725.58&12.6&83.92&0.12\\\bottomrule
	\end{tabular}}

\end{table}

\subsection{Final Remarks on the Results}\label{sub:remarks}

The consistently better results obtained by the SG-GRU for the GSOD dataset steem from the smoothness of the temperature GS with respect to the graph domain; SG-GRU relies on the assumption of bandlimited GSs. Therefore, SG-GRU is a promising approach to predict spatially smooth GSs. It is worth mentioning that the choice of the adjacency matrix is fundamental for a good performance, since it eventually defines the smoothness of the GSs. In the SeattleLoop dataset, which is not really smooth, the SG-GRU outperformed both the STGCN and the LSTM-based approaches when the sample size was small and the prediction time horizon was $30$ minutes, thus indicating that the SG-GRU can capture ST dependencies by taking the network frequency content into account. Moreover, SG-GRU has low computational cost and can be boosted with more recurrent or fully connected layers, when sufficient computational resources are available. 

\section{Conclusion}\label{sec:conclusion}
This work presented a new deep learning technique for jointly forecasting and interpolating network signals represented by graph signals. The proposed scheme embeds GSP tools in its basic learning-from-data unit (SG-GRU cell), thus merging model-based and deep learning approaches in a successful manner.  Indeed, the proposal is able to capture spatiotemporal correlations when the input signal comprises just a small sample of the entire network.  Additionally, the technique allows reliable predictions when input data is noisy or some values are missing by enforcing smoothness on the output signals. As future works, we envisage 
the use of the proposed SG-GRU as part of an anomaly detector in network signals, in which the anomalous sensors' measurements are characterized by large  deviations from the neighboring sensors.

\bibliographystyle{IEEEtran}
\balance
\bibliography{main.bib}

\end{document}